\documentclass[preprint]{elsarticle}

\usepackage{amssymb}
\usepackage{listings}

\usepackage{wrapfig}
\usepackage{subcaption}
\usepackage{lscape}
\usepackage{rotating}

\usepackage{xcolor}

\definecolor{codegreen}{rgb}{0,0.6,0}
\definecolor{codegray}{rgb}{0.5,0.5,0.5}
\definecolor{codepurple}{rgb}{0.58,0,0.82}
\definecolor{backcolour}{rgb}{0.95,0.95,0.92}

\lstdefinestyle{mystyle}{
    backgroundcolor=\color{backcolour}, ,   
    commentstyle=\color{codegreen},
    keywordstyle=\color{magenta},
    numberstyle=\tiny\color{codegray},
    stringstyle=\color{codepurple},
    basicstyle=\ttfamily\footnotesize,
    breakatwhitespace=false,         
    breaklines=true,                 
    captionpos=b,                    
    keepspaces=true,                 
    numbers=left,                    
    numbersep=5pt,                  
    showspaces=false,                
    showstringspaces=false,
    showtabs=false,                  
    tabsize=2
}
\usepackage{lineno,hyperref}

\journal{JSS-SI on Software Architecture and Artificial Intelligence}

\begin{document}

\begin{frontmatter}

\title{TOSCAdata: Modelling data pipeline applications in TOSCA}

\author[label1]{Chinmaya~Kumar~Dehury}
    \ead{chinmaya.dehury@ut.ee}
\author[label1]{Pelle~Jakovits}
    \ead{Jakovits@ut.ee}
\author[label2]{Satish~Narayana~Srirama  \corref{cor}}
    \cortext[cor]{Corresponding author}
    \ead{satish.srirama@uohyd.ac.in}
\author[label3]{Giorgos Giotis}
    \ead{g.giotis@atc.gr} 
\author[label1]{Gaurav Garg}
    \ead{gaurav.garg@ut.ee}

\address[label1]{Mobile \& Cloud Lab, Institute of Computer Science, University of Tartu, Tartu 50090, Estonia}
\address[label2]{School of Computer and Information Sciences, University of Hyderabad, Hyderabad 500 046, India}
\address[label3]{ Athens Technology Center S.A., Chalandri 15233, Athens, Greece}

\begin{abstract}

The serverless platform allows a customer to effectively use cloud resources and pay for the exact amount of used resources. A number of dedicated open source and commercial cloud data management tools are available to handle the massive amount of data. Such modern cloud data management tools are not enough matured to integrate the generic cloud application with the serverless platform due to the lack of mature and stable standards. One of the most popular and mature standards, TOSCA (Topology and Orchestration Specification for Cloud Applications), mainly focuses on application and service portability and automated management of the generic cloud application components. This paper proposes the extension of the TOSCA standard, \textit{TOSCAdata}, that focuses on the modeling of data pipeline-based cloud applications. Keeping the requirements of modern data pipeline cloud applications, TOSCAdata provides a number of TOSCA models that are independently deployable, schedulable, scalable, and re-usable, while effectively handling the flow and transformation of data in a pipeline manner. We also demonstrate the applicability of proposed TOSCAdata models by taking a web-based cloud application in the context of tourism promotion as a use case scenario. 

\end{abstract}
\end{frontmatter}
\pagenumbering{arabic}

\section{Introduction}\label{sec:intro}
Orchestration of cloud services is important for companies and institutions which need to design complex cloud-native applications or to migrate their existing services to the cloud. 
A number of orchestration solutions exist for clouds, but many of them are designed for specific platforms and introduce vendor lock-in, meaning it may be very difficult to migrate to other platforms when requirements change and the initially chosen technologies are no longer optimal \cite{opara-martins_critical_2016,casale_radon_2020}. Tools such as Chef \footnote{https://www.chef.io/}, Ansible\footnote{https://www.ansible.com/} and Puppet\footnote{https://puppet.com/} provide infrastructure-as-code (IaC) language to automate the installation and configuration of cloud applications, but are not simple to use for designing complex cloud systems consisting of tens or hundreds of components. 

The Topology and Orchestration Specification for Cloud Applications (TOSCA) \cite{matt_rutkowski_tosca_2020} language focuses on modeling the structure of cloud services to support their automation and orchestration. The developer can represent the structure of the services using node and relationship topology. The life-cycle of the cloud services can be managed automatically through different operations, such as create, start, configure, stop, etc. In addition, the relationship among those services can be managed automatically, for instance, which services to be configured before and after a relationship is created. 

One of the main advantages of the TOSCA standard is that it is platform-agnostic \cite{orazio2021torch}. Users can easily switch out node types representing cloud platform for other providers or even open source cloud software and enable the same cloud applications to be easily migrated between different vendors. It also is well compatible with different CI/CD technologies, which enable easier testing, re-deployment and re-engineering of applications expressed in TOSCA language \cite{artac2017devops}. 

\subsection{Motivation and goal}\label{sec:motiv_goal}
However, the building blocks of TOSCA modelling language mainly focus on automatic deployment and orchestration of generic cloud applications \cite{wild_tosca4qc_2020} and do not deal with controlling the flow of data inside such systems \cite{dehury_ccodamic_2020}. The challenge arises when data-intensive applications need to be designed and orchestrated. One of the crucial challenges in designing such cloud applications is the functionality to migrate the data in a multi-cloud system and integrate the serverless platform.  It requires not only orchestrating the infrastructure and software, but also to control the full life-cycle of data, from data ingestion to its migration, transformation, serverless platform integration, processing and storage.

With this motivation, in this work the TOSCA language has been extended and proposed as \textit{TOSCAdata} with the primary focus on efficiently handling the flow and processing of the data. Our goal is to take advantage of TOSCA capabilities to enable developers to rapidly model, develop, and deploy data pipeline applications, such as designing pipelines for migrating data between systems, tracking data versions, maintaining privacy and security, transforming data on-the-fly using serverless platform, processing with data analytics platform, fusing, merging the data, etc.

\subsection{Contributions}
The main contributions of TOSCAdata to TOSCA language are summarized as below:
\begin{itemize}
\item TOSCAdata extends the TOSCA standard with the ability to model the flow of data across cloud services.
\item It enables the data-pipelines-as-code pattern, allowing the automated deployment of data-pipelines (e.g., as part of CI/CD pipelines).
\item Reduces the development effort of designing data migration and processing services by providing reusable and freely combinable data pipeline blocks.
\item Reduces the effect of data lock-in by providing data pipeline blocks for different cloud providers (e.g. AWS, Azure, Google Cloud, OpenStack).
\item Ensures that data is encrypted while it is moved across cloud platforms over the internet (e.g., between on-premise and cloud, or in multi-cloud deployments).
\end{itemize}

The proposed \textit{TOSCAdata}, extends our previous work \cite{dehury_data_2020}, where the overall concept of how to model data pipeline applications with TOSCA language is presented with the intention to take advantage of the TOSCA features to model the flow of data in a multi-cloud system. 
The proposed extension of the TOSCA standard is also adopted in modelling of data pipeline applications in the RADON project \cite{casale_radon_2020}. The major extension made to our previous work \cite{dehury_data_2020} can be summarized below:
\begin{itemize}
    \item The methodology for orchestrating data pipeline services with the TOSCA standard is provided.
    \item A detailed description of each category of the data pipeline TOSCA node types, their specific functionalities, characteristics, implementation of all node types' life-cycle operations in Ansible, etc. are presented.
    \item After a thorough investigation it is concluded that previous research work does not provide a way to develop specific type of TOSCA node types that are suitable for standalone deployment. As a result, we have leveraged the TOSCAdata with a new \textit{Standalone} category of TOSCA node types.
    \item To ensure the deployability of the TOSCA service template, \textit{TOSCAdata Verifier} is designed.
    \item The extended work also focuses on design and development of the essential features, such as event and cron-based scheduling, of TOSCAdata that may be required while migrating the data across multi-cloud environments.
    \item The application of TOSCAdata in real world tourism promotion application is explained and demonstrated using the Viarota application.
    \item In addition to a real-world application, TOSCAdata is also demonstrated by implementing an image data migration and processing application as described in Section \ref{sec:usecase:myDemo}. The demonstration example shows the movement across four different private and public clouds and the integration of serverless platforms to process those data.
\end{itemize}

Further to demonstrate the purpose and capabilities of the proposed work that extends the TOSCA language, we have applied the extended works on Viarota \cite{viarota}, a mobile and web-based cloud application in the context of tourism promotion, as a use case. In this use case, the data needs to be synchronised between the storage system available in three different public clouds: AWS cloud, Google cloud and Azure cloud. A detailed description, including the shortcoming of existing TOSCA, benefits of adopting data pipeline nodes in the use case implementation, is presented in Section \ref{sec:usecase}.


The rest of the paper is organized as follows: 
Section \ref{sec:background} presents the technical background and the related works about the data pipeline and TOSCA.
Section \ref{sec:sol} outlines the methodology of using TOSCA for modelling data pipeline services. 
Section \ref{sec:nodetypes} describes the extensions introduced to TOSCA language.
Section \ref{sec:usecase} explains the use case scenarios followed by the concluding remarks in the Section \ref{sec:concl}. 


\section{Background and Related works}\label{sec:background}
This section discusses the technical background on TOSCA and the recent related works on data pipeline modeling.

\subsection{Data pipeline-based cloud applications}
Data pipeline mainly refers to automatizing the basic three operations: Extract, Transform, and Load in a pipeline manner. A data pipeline application consists of a large number of independent pipeline blocks \footnote{Henceforth, the term \emph{pipeline block} and \emph{pipeline} are used interchangeably.} connected sequentially. The output of one pipeline block is the input for another pipeline block. Each block is composed of one or more microservices, serverless functions, or self-contained applications designed to perform a specific task for data transformation, storage, and processing. Such blocks are freely composable, portable, and reusable; designed in such a way that they are independently deployable, schedulable, and scalable in the cloud environment. For example, a block could be to read the list of files present in a remote file server. Simultaneously, another block could be designed to get the files one after another based on the list. Another example of a pipeline block could be to read the images from the AWS S3 bucket and store them in a local directory. 
Between each pair of pipeline blocks, temporary buffer storage is used to balance the throughput of pipeline blocks at both ends. 

Pervaiz et al. \cite{pervaiz_examining_2019} investigated the challenges that are faced in the phase of data cleaning and data processing in development data. It is observed that, even with modern technology, it is difficult to maintain data consistency during the transition of data between data collection, data cleansing, and data analytics. The data pipeline approach is also applied in the evaluation of supervised and unsupervised machine learning algorithms in smart transportation \cite{howard_distributed_2018} systems. The algorithms are designed to execute on a distributed system, employing different AWS cloud services, such as S3 bucket and MongoDB for data storage, EC2 and EMR cluster for providing computational resources, and Spark for big data processing platform.

One of the main hurdles in cloud adoption of data-intensive applications is the absence of mature data management solution that addresses vendor lock-in issue \cite{casale_radon_2020, opara-martins_critical_2016} despite of several open-source and commercial data management solutions available in the market, such as Apache NiFi \cite{nifi}, AWS data pipeline \cite{noauthor_aws_dp_devguide}, Google dataflow \cite{google_dataflow}, Azure data factory \cite{Azure_azure}, etc. Mainly one open-source (Apache NiFi) and one commercial data management solutions (AWS data pipeline) are taken into account in this work.

Apache NiFi \cite{nifi}, an open-source data management solution, focuses on smooth and efficient flow and processing of data through a large number of components called processors. To connect one processor with others, input ports and output ports are used. Apache NiFi internally creates the intermediate temporary storage and implements a queuing system between adjacent connected processors. The processors are developed for different purposes, such as processors for interacting with AWS storage, Google Storage, Microsoft storage system, different serverless platforms, the transformation of text data, invoking local system commands, etc.

On the other hand, AWS data pipeline (DP) service \cite{noauthor_aws_dp_devguide} is a service that especially focuses on the flow and processing of data within different services provided by AWS cloud. Data in the AWS cloud can be transformed and moved from one AWS cloud service to another using different activities, such as CopyActivity, PigActivity, SqlActivity, etc. These activities act as basic pipeline components/blocks that apply certain operations on the data nodes. Different data nodes that AWS DP supports are S3DataNode, SqlDataNode, DynamoDBDataNode, and RedshiftDataNode. In general, it is observed that, AWS DP does not provide enough stable mechanisms for a smooth and efficient flow of data in and out AWS cloud platform. To move the data in and out of the AWS cloud ecosystem, user needs to issue a shell command using ShellActivity data pipeline. There is no specific data pipeline to migrate the data from/to any AWS storage service \cite{noauthor_aws_dp_devguide}. This introduces a data lock-in issue while developing large data-intensive cloud applications. 

The security breach is another major issue that may bring huge monetary loss to the business \cite{byrne_development_2020}. This may occur due to the lack of platform-specific expertise and data mishandling in the public cloud.
Another research challenge in handling the data in a multi-cloud data pipeline platform is the recovery of the data from its failure, as this involves different storage units from different cloud providers, as discussed in \cite{wang_system_2019}. Another reason that makes this research challenge more complex is due to the diverse backup and recovery services provided by multiple cloud platforms and the lack of co-operations between those services.

\subsection{TOSCA language for modeling cloud applications}\label{sec:TOSCABackground}
\begin{figure}[ht]
 \centering
 \includegraphics[width=\linewidth]{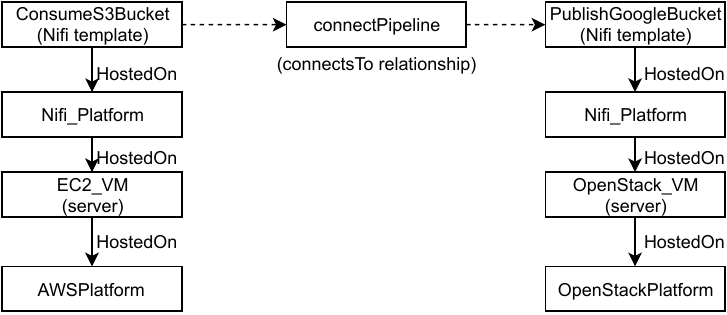}
 \caption{An example of TOSCA topology.}
 \label{fig:toscaTopologyExample}
\end{figure}

On the other hand, Topology and Orchestration Specification for Cloud Applications (TOSCA) \cite{matt_rutkowski_tosca_2020} is a recently developed OASIS standard focusing on the portability and interoperability of cloud-based applications. In TOSCA language, a service blueprint describes the structure of the whole cloud application along with the management aspect (i.e., deployment, operation, termination) of each component. The service blueprint consists of a set of \textit{nodes} (to represent software components) and \textit{edges} (to represent the relationship among software components). A data pipeline block is called a node in TOSCA language. These set of nodes and the edges form a \textit{topology template}. Here each node represents a single independent or dependent software component, and an edge represents the relationship (i.e. \emph{HostedOn}, \emph{ConnectsTo}, \emph{DependsOn}, etc. \cite{OpenTOSCA,6188582}) between two software components. An example of the TOSCA topology is presented in Figure \ref{fig:toscaTopologyExample} that copies the data from AWS S3 bucket to Google cloud Storage bucket using  NiFi. The topology consists of eight nodes/pipeline blocks: \textit{AWSPlatform}, \textit{OpenStackPlatform}, \textit{EC2\_VM}, \textit{OpenStack\_VM}, two \textit{Nifi\_Platform} nodes, \textit{ConsumeS3Bucket}, and \textit{PublishGoogleBucket}. This also demonstrates \textit{HostedOn} and \textit{ConnectsTo} relationships. TOSCA service blueprints follow the YAML syntax, and all the related definitions and artifacts are encapsulated in a CSAR (Cloud Service Archive) file, which is a standardized packaging format.

TOSCA Simple Profile \cite{matt_rutkowski_tosca_2020} provides a set of normative node types, relationship types, capabilities types, data types, different types of interfaces, etc. Such normative types can be used to extend and define desired nodes, relationships, capabilities types based on the requirement of application modelling. Each node type definition consists of a set of properties, attributes, requirements, capabilities, implementations, etc. Some examples of normative node types are \textit{Compute}, \textit{SoftwareComponent}, \textit{WebServer}, \textit{WebApplication}, \textit{DBMS}, \textit{Database}, \textit{ObjectStorage}, etc. 

\textit{Requirements} and \textit{capabilities} of a TOSCA node type are counterpart to each other. For any relationship between two nodes, first node should have the requirement to connect to the second node and similarly the second node should have capability to accept the connection from the first node. The \textit{requirements} of a node type describe the needs of that node, which can be in terms of hosting environment, computing or storage resource, or any software component. On the other hand, the required node type should have the corresponding capability to fulfill the demand of counterpart TOSCA node. For instance, in Figure \ref{fig:toscaTopologyExample}, the requirement of \textit{EC2\_VM} is a AWS hosting platform of type \textit{AWSPlatform}. On the other hand, \textit{AWSPlatform} should have the capability to host a \textit{EC2\_VM} type TOSCA node. Similarly, the requirements of \textit{ConsumeS3Bucket} are \textit{Nifi\_Platform} and \textit{PublishGoogleBucket}. At the same, \textit{Nifi\_Platform} must have the capability to host \textit{ConsumeS3Bucket} TOSCA node and \textit{PublishGoogleBucket} must have the capability to accept connection from \textit{ConsumeS3Bucket}.

Each node type can have one or more capabilities, such as \textit{Compute}, \textit{Network}, \textit{Storage}, \textit{Container}, \textit{Endpoint}, etc. On the contrary, a node type definition can have more than one requirements depending on the node type this needs to be connected to. When two node types are connected, a relationship between them needs to be defined. The normative relationship types provided by TOSCA Simple Profile are \textit{HostedOn}, \textit{ConnectsTo}, \textit{DependsOn}, \textit{AttachesTo}, and \textit{RoutesTo}.

The lifecycle of each node is implemented using \textit{create}, \textit{configure}, \textit{start}, \textit{stop}, and \textit{delete} operations. Depending on the TOSCA orchestration platform, the implementation file(s) for each operation can be provided using Ansible script, Python script, or other scripting languages. A number of commercial and open-source TOSCA orchestration platforms are developed, such as Cloudify \cite{noauthor_cloudify_nodate}, xopera \cite{xopera}, etc. It is to be noted that TOSCA only provides a high-level description of cloud applications. The high-level description contains the properties, attributes, requirements, and capabilities of each respective component. Further, with each component, a standard interface is attached that describes the lifecycle of the component. It is the responsibility of the orchestrator to understand and implement each node along with the associated relationships. The orchestrator provides the necessary runtime environment to invoke the implementation file for each lifecycle operation. An implementation file provides information that is enough to make automatic deployment and un-deployment of the applications, provisioning of the resources, and manage the lifecycle of the application, etc. \cite{kopp_winerymodeling_2013}.

\subsection{Literature survey on TOSCA}
Along with commercial and non-profit organisations, TOSCA has been widely adopted in other technologies, such as the Internet of Things (IoT) \cite{da_silva_opentosca_2016} \cite{franco_da_silva_internet_2017} \cite{li_towards_2013}, Network Function Virtualization (NFV) \cite{antonenko_c2_2017} \cite{brito_service_2017, hung_orchestration_2017}, quantum computing \cite{wild_tosca4qc_2020}, fog and mobile edge computing \cite{brito_service_2017}. TOSCA is mainly designed and developed to improve the portability and interoperability of the cloud technology \cite{di_martino_ontology_2020}. This has also been extended to support the scaling \cite{cankar_auto-scaling_2020}, load balancing, monitoring, and other aspects of cloud computing. Dehury et al in \cite{dehury_ccodamic_2020,dehury_data_2020} developed the TOSCA profile for serverless and data pipeline-based cloud applications. 

The adoption of TOSCA Simple Profile in the field of IoT can be found in early 2013 by Li et al. \cite{li_towards_2013}, where a set of node type definitions, such as \textit{Controller}, \textit{Gateway}, and \textit{Driver}, along with the modelling of required relationship types are derived keeping the requirement of building automation system in mind. Similarly, Silva et al. \cite{franco_da_silva_internet_2017} adopt the TOSCA Simple Profile for automatic deployment and setup of IoT environments. This addresses the major challenges due to the heterogeneous nature of all the components, such as sensors, deployment environments, actuators, etc., of the IoT environment. The authors here proposed TOSCA profiles for defining and setting up hardware components, deployment of of middlewares, and deployment of IoT applications atop the IoT middlewares. For exchange of messages among the IoT devices, authors in \cite{da_silva_opentosca_2016} propose the required TOSCA profiles for deployment and configuration of MQTT brokers. 

TOSCA is further extended to handle several cloud deployment challenges leading to its multi-dimensional extensions. Kehrer et. al. \cite{kehrer2018tosca} took advantage of the TOSCA to address the lock-in problem in the container management system. Authors propose a two-phase deployment method for Mesos to integrate the orchestration of cloud service based on TOSCA and their automation using the container-based artifacts. However, this extension of TOSCA entirely focuses on the orchestration and automation aspect of the cloud services. Further, on efficient management of complex cloud services across the heterogeneous platform,  Brogi et al. \cite{brogi2018tosker} extended the TOSCA standard for the orchestration and lifecycle management of the Docker-based software components. Furthermore, the TOSCA standard is also extended to the Kubernetes for the cloud-based enterprise applications \cite{bogo2020component}. The above extensions mainly focus on the cloud application's deployability and not the data migration and processing.

To handle the above issue, TOSCA is also adopted to design and orchestrate big data architecture and services \cite{Guerriero2016BigData}. Keeping that in focus, the authors have developed a set of TOSCA profiles for the data source, storage, computation of data, data specification, etc. However, the developed TOSCA models are most suitable for big-data application and don't take the pipeline characteristics of data migration and processing. Further, there is no focus on the integration of the serverless platform with the big data application. To handle the data exchange in the IoT environment and their customization and provisioning of the complex event processing (CEP), Silva et al. \cite{da2018customization} took the capability of TOSCA standard and proposed an approach that efficiently handles the CEP systems, especially in an IoT environment. However, the proposed system does not take the lifecycle of the data flow into consideration.

Broadening the applicability of the TOSCA from just cloud service management, Tsagkaropoulos et al. \cite{tsagkaropoulos2021extending} extended the TOSCA standard for providing support to edge and fog deployment of the services. The authors have developed a set of TOSCA node types, policies, relationships, capabilities, etc suitable for deployment of software components in fog environment taking the resource constraints into consideration.    

\section{TOSCAdata: Data pipeline modelling using TOSCA}\label{sec:sol}
This section presents the proposed \textit{TOSCAdata}, the extended version of the TOSCA standard for cloud-based data pipeline applications. Considering the motivation and goal, as mentioned in Section \ref{sec:motiv_goal}, this section provides a detailed explanation on the features, capabilities and functionalities that \textit{TOSCAdata} facilitates to the TOSCA standard. The extension to TOSCA standard includes design and development of essential TOSCA profiles for consuming the data from the source, publishing the result to sink and processing the intermediate data by enabling serverless platform integration.  

As discussed before, data pipeline applications mainly focus on extraction, transformation, and migration of the data. Such applications are composed of a large number of pipeline blocks that focus on different data-related operations. Further, each block is independent, freely composable, portable, and reusable. This section discusses on methodology to develop such independent pipeline blocks using TOSCA.


\begin{figure}[h]
 \centering
 \includegraphics[width=\linewidth]{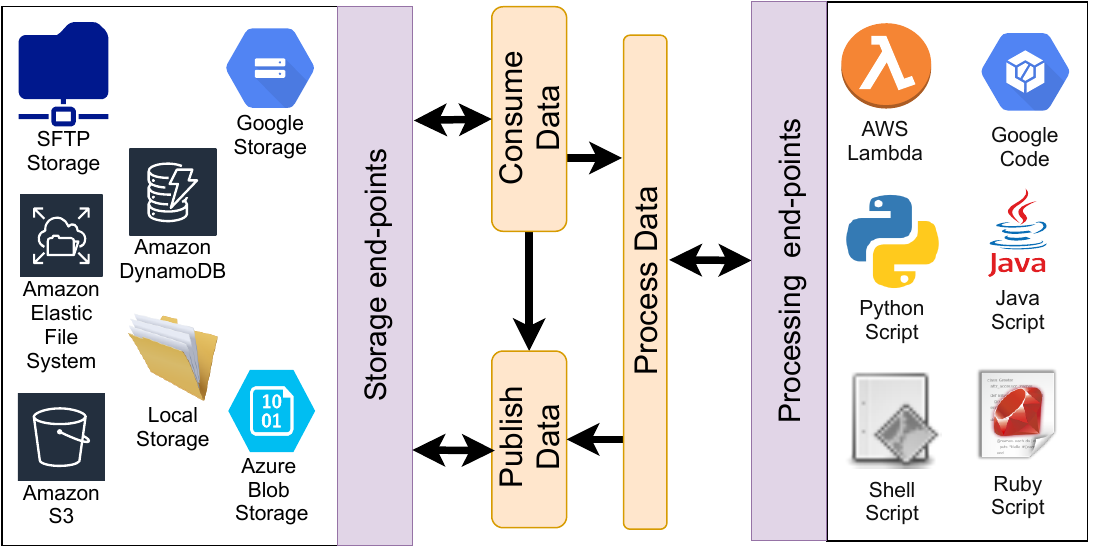}
 \caption{Basic building blocks of data-focused cloud services.}
 \label{fig:BasicBuildingBlocks}
\end{figure}
Any data flow-based cloud service can be generalized as a combination of three basic building blocks: (a) consuming the data, (b) processing the consumed data, and (c) publishing the processed or consumed data, as shown in Figure \ref{fig:BasicBuildingBlocks}.  \textit{Consuming} and \textit{Publishing} data refer to reading and writing the data from/to local/remote storage end-point. The local storage end-point refers to the local directory structure of the same host machine where the service component is running. The remote storage end-point refers to the storage bucket provided by other CSPs, such as Amazon S3 bucket, Google cloud storage bucket, Azure blob storage, etc. The remote storage end-point can also be an SFTP server. On the other hand, \textit{Processing} of the consumed data can be achieved by invoking the remote serverless functions, such as Amazon Lambda, Google function, Azure function, OpenFaaS function, etc. The processing can also be done in the local machine by using shell script, Python script, Ruby script, or any other programming language script.

\begin{sidewaysfigure}
    \centering
    \includegraphics[width=0.98\textwidth, scale=0.9]{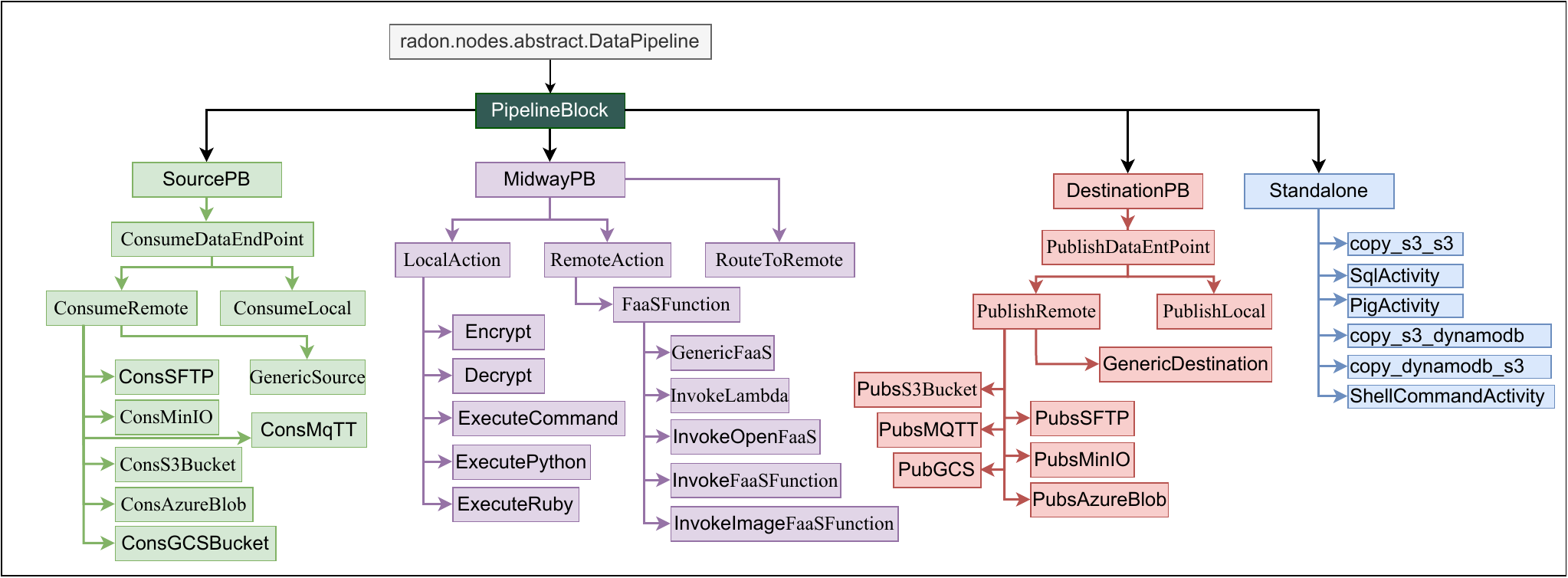}
    \caption{TOSCA data pipeline models hierarchy.}
    \label{fig:ModelHrchy}
\end{sidewaysfigure}

Based on the above basic data pipeline building blocks, the proposed model hierarchy provides a set of TOSCA-based data pipeline models to fulfill the requirements of each building blocks that are used to develop modern data flow-based cloud services. The TOSCA models are mainly classified into three categories: \textit{SourcePB}, \textit{MidwayPB}, \textit{DestinationPB}, which are derived from an upper level of TOSCA node type \textit{PipelineBlock} (PB), as shown in Figure \ref{fig:ModelHrchy}. The detailed description of those TOSCA node types are given in Section \ref{sec:nodetypes}.

As shown in Figure \ref{fig:ModelHrchy}, \textit{PipelineBlock} (PB) TOSCA node type is derived from the abstract node type, \textit{radon.nodes.abstract.DataPipeline}. The definition of \textit{PipelineBlock} TOSCA node type contains the definitions of properties, attributes, and other artifacts that are common to all the derived pipeline blocks, as in Listing \ref{code:PipelineBlock}. 

\subsection{TOSCAdata properties \& attributes}
The properties and attributes are specific to each TOSCA DP node types. However, it is found that some properties and attributes can be common to all the DP node types, such as the name of the pipeline block, its scheduling strategy, etc., and are defined in \textit{PipelineBlock} node type, as shown in Listing \ref{code:PipelineBlock}. Some of the common properties are the \textit{name} of the pipeline, which is of \textit{string} data type, as in Line-15. For each pipeline, a user needs to mention the name of the pipeline in the service template. It is possible to assign two pipelines with the same name without any runtime conflict that may occur during the orchestration of the service template. This is due to the fact that each pipeline is assigned with a unique ID to the \textit{id} attribute (say in Listing \ref{code:PipelineBlock}, Line 6) of the corresponding pipeline, and such unique IDs are used for future reference. The \textit{id} of a pipeline is mainly generated by the underlined pipeline technology used by the implementation file. If the pipeline is based on Apache NiFi, the NiFi engine will generate a unique ID. However, if the pipeline is based on ADP, the unique id would be generated by the corresponding AWS service \cite{noauthor_aws_dp_devguide}. Hence, the generation of unique IDs of the pipelines is very specific to and dependent on the underlined pipeline technology.

\begin{lstlisting}[language=xml, label=code:PipelineBlock, caption=PipelineBlock TOSCA node type]
 
tosca_definitions_version: tosca_simple_yaml_1_3
node_types:
  radon.nodes.datapipeline.PipelineBlock:
    derived_from: radon.nodes.abstract.DataPipeline
    attributes:
      id:
        type: string
    properties:
      schedulingStrategy:
        type: string
        default: "EVENT_DRIVEN"
      schedulingPeriodCRON:
        type: string
        default: "* * * * * ?"
      name:
        type: string
\end{lstlisting}

The \textit{PipelineBlock} in Listing \ref{code:PipelineBlock} also contains the definition of \textit{schedulingStrategy}, Line 9, and \textit{schedulingPeriodCRON}, Line 12. These two common properties allow the developers to schedule the pipelines that are triggered based on specific events or time. Time-driven scheduling is implemented using the CRON scheduler. The detailed description of the scheduling properties of the pipelines and their implementation are described in the Sub-section \ref{sec:cap_features}. 

\subsection{TOSCAdata requirements, relationships \& capabilities}
\begin{figure}[h]
 \centering
 \includegraphics[width=0.9\linewidth]{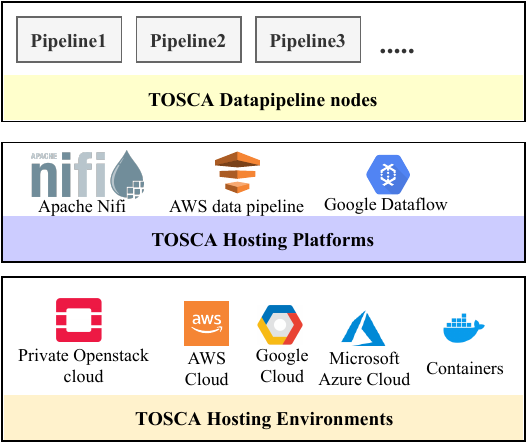}
 \caption{Hosting hierarchy of TOSCA pipelines, platforms, and environments.}
 \label{fig:HostingHrchy}
\end{figure}

The developed TOSCA data pipeline nodes/blocks have three requirements: (a) requirement to connect to a local pipeline (\textit{connectToPipeline}), (b) requirement to connect to a remote pipeline (\textit{connectToPipelineRemote}), and (c) requirement of a hosting environment (\textit{host}), based on the functionality of the node. \textit{connectToPipeline} and \textit{connectToPipelineRemote} indicates the requirement to connect to other pipeline, whereas \textit{host} indicates the requirement of an hosting platform, which in turn requires an hosting environment, as shown in Figure \ref{fig:HostingHrchy}.
The relationship between two local or remote pipelines can be of type \textit{ConnectNifiLocal} or \textit{ConnectNifiRemote}, which are derived from the \textit{tosca.relationships.ConnectsTo} normative relationship type available in TOSCA simple profile v1.3 \footnote{https://docs.oasis-open.org/tosca/TOSCA-Simple-Profile-YAML/v1.3/os/TOSCA-Simple-Profile-YAML-v1.3-os.html}. Both the relationship types can be differentiated by means of their implementations only. Two NiFi-based pipelines are connected by connecting the output port of the source pipeline to the input port of the destination pipeline. However, if the pipelines are on different host machines, we create a Remote Process Group (RPG) on the host machine where the source pipeline is deployed. The implementation file for performing this job is created using Ansible. 

Further, as mentioned above, that the developed DP node types require a hosting environment. The current development of TOSCA data pipeline node types supports the Apache NiFi and AWS data pipeline as the hosting platforms. However, depending upon the node type, the compatible hosting platform needs to be chosen. Apache NiFi hosting platforms can be deployed on a virtual machine or a container on either a private cloud or public cloud, as shown in Figure \ref{fig:HostingHrchy}. The TOSCA node type for Apache NiFi, \textit{NiFi}\footnote{\url{https://github.com/radon-h2020/radon-particles/nodetypes/radon.nodes.nifi/Nifi}}, hosting platform is given in Listing \ref{code:nifi_platform}. \textit{NiFi} node type has the properties \textit{port} number (to access NiFi's web interface), in Line 5-1, and the \textit{component\_version} (to specify the version of the Apache NiFi), in Line 8-9. This hosting platform has the capability to host \textit{UNBOUNDED} number of NiFi-based pipelines, in Line 17-20, which indicates no limit to the number of data pipeline nodes that can be hosted on. Further, this hosting platform requires a hosting environment, as in Line 11-15, with normative \textit{tosca.capabilities.Compute} capability. As shown in Figure \ref{fig:HostingHrchy}, the hosting environment can be OpenStack-based private cloud or public clouds, such as AWS cloud, Google cloud, Microsoft Azure cloud, etc.

\begin{lstlisting}[language=xml, label=code:nifi_platform, caption= ]
node_types:
  radon.nodes.nifi.Nifi:
    derived_from: tosca.nodes.SoftwareComponent
    properties:
      port:
        type: string
        default: 8080
      component_version:
        type: string
    requirements:
      - host:
          capability: tosca.capabilities.Compute
          node: tosca.nodes.Compute
          relationship: tosca.relationships.HostedOn
          occurrences: [ 1, 1 ]
    capabilities:
      host:
        occurrences: [ 1, UNBOUNDED ]
        valid_source_types: [ radon.nodes.abstract.DataPipeline ]
        type: tosca.capabilities.Container
\end{lstlisting}

The developed TOSCA DP node types have the capabilities to accept the data through the incoming connection from other pipelines to either publish or process the data. For this, ConnectToPipeline capability type is developed, which is derived from \textit{tosca.capabilities.Endpoint} TOSCA normative capability type, as shown in Listing \ref{code:capability_type}. The data may arrive from one or more local or remote pipelines. 

\begin{lstlisting}[language=xml, label=code:capability_type caption= TOSCA data pipeline capability type.]
tosca_definitions_version: tosca_simple_yaml_1_3
capability_types:
  radon.capabilities.datapipeline.ConnectToPipeline:
    derived_from: tosca.capabilities.Endpoint
    metadata:
      targetNamespace: "radon.capabilities.datapipeline"
      abstract: "false"
      final: "false"
\end{lstlisting}


\subsection{TOSCAdata functionalities}\label{sec:cap_features}
With the current development of TOSCAdata, a number of functionalities and features are provided. Some of the major functionalities are supported for scheduling the pipeline-based on event-driven or time-driven data flow across multiple private and public clouds by addressing data lock-in issue, and incorporation of the serverless platform, as discussed below.

\subsubsection{Scheduling of data pipelines}
The scheduling definition is implemented in \textit{PipelineBlock} TOSCA node, as shown in Listing \ref{code:PipelineBlock}, Line 9-14. Two properties \textit{schedulingStrategy} (Listing \ref{code:PipelineBlock}, Line 9-11) and \textit{schedulingPeriodCRON} (Listing \ref{code:PipelineBlock}, Line 12-14) are introduced. \textit{schedulingStrategy} property allows the user to decide whether the pipeline should be triggered based on events or a specific time. The default scheduling strategy is set to be event-driven. The scheduling strategy of Apache NiFi is used to schedule the NiFi-based pipelines. For timer driven, users need to provide time interval using CRON syntax into \textit{schedulingPeriodCRON} properties. However, for the AWS related TOSCA pipelines (as discussed in detail in later sections), only CRON-based scheduling is supported, and hence the above two properties \textit{schedulingStrategy} and \textit{schedulingPeriodCRON} are not made available to AWS related TOSCA pipelines.

\subsubsection{Cross cloud data flow}
With the developed set of TOSCA DP node types, it is easy to allow the data movement across different clouds. This is demonstrated in the Use case section (Section \ref{sec:usecase}), where, with the minimal design time effort, the data are migrated from one cloud storage (e.g. AWS S3 Bucket) to the other (e.g. Google Cloud Storage). TOSCAdata provides an essential set of models to consume and publish the data from and to a number of storage systems. As it can be seen in Figure \ref{fig:ModelHrchy}, a storage system can be in local directory structure (\textit{ConsumeLocal} node type), or from AWS S3 bucket and DynamoDB (\textit{ConsS3Bucket}, \textit{PubsS3Bucket} \& \textit{copy\_dynamodb\_s3} node types), Google cloud storage (\textit{ConsGCSBucket} \& \textit{PubGCS} node types), Azure blob storage (\textit{ConsAzureBlob} \& \textit{PubsAzureBlob} node types), on-premise MinIO server (\textit{ConsMinIO} \& \textit{PubsMinIO} node types), etc. It is also possible to consume the data from edge devices or resource constraint devices over light-weight MQTT protocol (\textit{ConsMqTT} \& \textit{PubsMQTT} node types) and the data external sources over SFTP protocol (\textit{ConsSFTP} \& \textit{PubsSFTP} node types). TOSCAdata also provides a set of node types to integrate the serverless platform that enables the data-on-fly to be processed by user-defined FaaS functions. Some of such node types are: \textit{InvokeLambda} to invoke AWS lambda function, \textit{InvokeOpenFaaS} to invoke a FaaS function deployed in OpenFaaS environment, \textit{InvokeFaaSFunction} to invoke any FaaS function over HTTP request, and \textit{InvokeImageFaaSFunction} to invoke generic FaaS function only with image data as arguments. The detailed descriptions are provided in Section \ref{sec:nodetypes}.

\subsubsection{TOSCAdata DP Verifier}
In the design time, the TOSCA service blueprint may contain inconsistency/bugs related to inappropriate connections among nodes. The orchestration tool (such as TOSCA orchestration or xopera \cite{xopera}) may generate unexpected errors during the deployment of nodes, or the user may get undesired result during the runtime if such erroneous TOSCA service templates are used for the deployment of the cloud services. To overcome such issues, the \textit{TOSCAdata Verifier} \footnote{\url{https://github.com/radon-h2020/radon-datapipeline-plugin}} tool is developed that ensures the workability of the TOSCA service template. The verifier parses through the nodes defined in the service blueprint and verifies the relationship, capabilities, and requirements. In case any mismatch is detected, it resolves the error and generates a verified service blueprint. 

\begin{figure*}[h]
 \centering
 \includegraphics[width=0.8\linewidth]{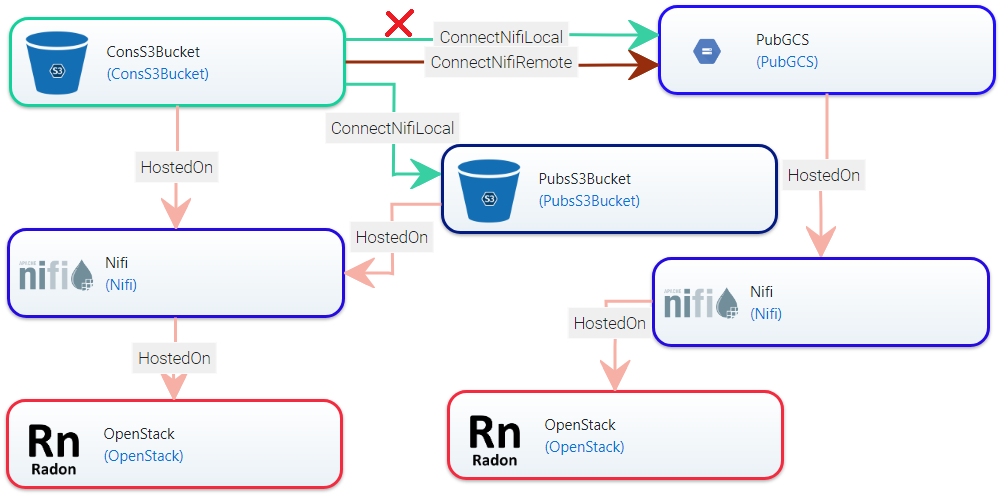}
 \caption{An example of erroneous TOSCA service Template.}
 \label{fig:erroneous_ServiceTemplate}
\end{figure*}

Figure \ref{fig:erroneous_ServiceTemplate} shows an example of an erroneous TOSCA service template. In this design, the node \textit{ConsS3Bucket} is hosted on one \textit{NiFi}, while node \textit{PubGCS} is hosted on another \textit{NiFi} platform on different virtual machine. The defined connection type between them should be remote \textit{ConnectNifiRemote}. However, it may happen that there exist two connections of different relationship types between the same pair of TOSCA nodes, as shown in Figure \ref{fig:erroneous_ServiceTemplate}. In this figure two different connections: \textit{ConnectNifiRemote} and \textit{ConnectNifiLocal}, exist between \textit{ConsS3Bucket} and \textit{PubGCS}. This is a good example of an inconsistent TOSCA service template, which causes the TOSCA orchestrator to generate errors while trying to deploy the nodes. Such potential errors can be eliminated by parsing the blueprint through the TOSCA DP verifier.

Apart from finding and fixing the design-time errors in the service template, the DP verifier also checks for the properties of \textit{Encrypt} and \textit{Decrypt} node types. These two node types are designed to provide a way to encrypt the content while sending the data from one machine to another and decrypt the received encrypted content. While doing so, it is essential to ensure that both the nodes at different machines have the same passphrase for encryption and decryption purposes. In case of the presence of such passphrase mismatch, the verifier will generate a new passphrase for both the nodes pair and accordingly update the service template. The DP verifier also ensures that for each \textit{Encrypt} type node in the service template, there exists a \textit{Decrypt} type node and vise versa. 

The current version of the \textit{TOSCAdata Verifier} can verify and update the relationships among the data pipelines and the encryption configuration for the secure transmission of the data. However, the verifier can further be extended to handle a number of runtime configuration of pipelines. Some of the features that are under development and will be incorporated into the \textit{TOSCAdata Verifier} are: ensuring that the right set of artefacts are provided to each data pipelines, the TOSCA definition for each pipeline is available in the service template, ensuring that the properties values are of valid type, etc.

\section{TOSCAdata models}\label{sec:nodetypes}
This section discussed the developed models that are introduced in Section \ref{sec:sol}, for \textit{TOSCAdata} using the TOSCA language. For the implementation of the lifecycle of developed TOSCA models, Ansible \footnote{https://docs.ansible.com/} is used, which provides a set of modules for software provisioning, configuration management, and application deployment. As discussed before, the developed TOSCA data pipeline nodes are based on Apache NiFi and AWS data pipeline. To be more precise, the node types that are developed under \textit{SourcePB}, \textit{MidwayPB}, and \textit{DestinationPB} categories to consume, process and publish the data, respectively, are based on the Apache Nifi technology. All the node types that are developed on \textit{Standalone} data pipeline category are based on the AWS data pipeline technology. During the orchestration of NiFi-based TOSCA nodes, the corresponding pre-designed NiFi template is uploaded to the NiFi hosting platform. Such NiFi templates are exported in XML format and may compose of one or more NiFi processors. The corresponding REST requests are issues for uploading the NiFi template to the remote NiFi platform, deployment of the NiFi templates, and activating the processors. \textit{uri} \footnote{\url{https://docs.ansible.com/ansible/latest/collections/ansible/builtin/uri_module.html}} Ansible module is used to issue those REST requests.

\subsection{SourcePB}
This subsection discusses a specific category of the TOSCA node types (including their requirements and capabilities) specifically developed to consume the data from several external sources, such as FTP server, Google storage bucket, AWS S3 bucket, MinIO servers, MQTT broker, Azure storage system etc.

For any data flow-based cloud application consuming the data from any local or remote location is one of the basic building blocks. To address this \textit{SourcePB}\footnote{\url{https://github.com/radon-h2020/radon-particles/nodetypes/radon.nodes.datapipeline.source}} TOSCA node type is defined that is derived from \textit{PipelineBlock} TOSCA node type. \textit{SourcePB} node type is further used to derive \textit{ConsumeDataEndPoint} node type. \textit{SourcePB} node type defines the \textit{requirements} that are common to any type of data end-point. The data or storage endpoint may refer to SFTP storage server, Google storage, Amazon Elastic File System, Amazon S3, Azure storage or a local storage, as shown in Figure \ref{fig:BasicBuildingBlocks}. The TOSCA nodes that are of type \textit{SourcePB} or of type derived from \textit{SourcePB} may have three requirements, as shown in Listing \ref{code:sourcepb_requirements},: (a) \textit{connectToPipeline} requirement, in Line 2-6, (b) \textit{connectToPipelineRemote} requirement, in Line 7-11, and (c) \textit{host}, in Line 12-16.

\begin{lstlisting}[language=xml,label=code:sourcepb_requirements, caption= List of requirements for \textit{SourcePB} and all derived node types]
requirements:
  - connectToPipeline:
      capability: radon.capabilities.datapipeline.ConnectToPipeline
      node: radon.nodes.abstract.DataPipeline
      relationship: radon.relationships.datapipeline.ConnectNifiLocal
      occurrences: [ 1, UNBOUNDED ]
  - connectToPipelineRemote:
      capability: radon.capabilities.datapipeline.ConnectToPipeline
      node: radon.nodes.abstract.DataPipeline
      relationship: radon.relationships.datapipeline.ConnectNifiRemote
      occurrences: [ 1, UNBOUNDED ]
  - host:
      capability: tosca.capabilities.Container
      node: radon.nodes.nifi.Nifi
      relationship: tosca.relationships.HostedOn
      occurrences: [ 1, 1 ]
\end{lstlisting}

Such TOSCA nodes require to send the consumed data to other pipelines of type, either \textit{MidwayPB} or \textit{DestinationPB}. \textit{connectToPipeline} and \textit{connectToPipelineRemote} requirements are used to connect to the other end of the pipelines deployed on the same and different virtual machines or the containers, respectively. From its functionalities, it is obvious that such TOSCA nodes do not need any capabilities to receive the data from other pipelines. Hence, all the TOSCA nodes of type \textit{SourcePB} can not send the data to another \textit{SourcePB} TOSCA node. In other words, the consumed data by \textit{SourcePB} TOSCA nodes can not be treated as the data source for another \textit{SourcePB} TOSCA node.

All the \textit{SourcePB}-derived TOSCA node types are based on Apache NiFi data management technology. Hence such TOSCA nodes require an Apache NiFi hosting platform. As discussed above, the Apache NiFi hosting platform can be deployed either in the AWS cloud platform or Openstack-based private cloud platform. 

Based on the storage location, two TOSCA node types are created: \textit{ConsumeRemote} and \textit{ConsumeLocal}. \textit{ConsumeRemote} TOSCA node type is further used to derived a number of other TOSCA node types for consuming data from remote storage servers. When the source of data is other than the local machine where the pipeline is deployed, \textit{ConsumeRemote} TOSCA node type is used. For different data sources, \textit{ConsFTP}, \textit{ConsGCSBucket}, \textit{ConsS3Bucket} etc. TOSCA node types are developed. 

\textit{ConsFTP} is used to consume the data from a FTP server. Similarly, \textit{ConsGCSBucket} (in Listing \ref{code:ConsGCSBucket_properties}) and \textit{ConsS3Bucket} (in Line \ref{code:ConsS3Bucket_properties}) TOSCA node types are developed to consume the data from Google cloud storage bucket and AWS S3 bucket. For \textit{ConsGCSBucket} TOSCA node type has three main properties: \textit{bucket} (Listing \ref{code:ConsGCSBucket_properties}, Line 5-8) for providing the name of the bucket, \textit{project\_id} (Line 9-11) for the id of the project, and \textit{credential\_JSON\_file} (Line 12-14) for path the file that contains login credentials, as shown in Listing \ref{code:ConsGCSBucket_properties}. Similarly, \textit{ConsS3Bucket} TOSCA node type has three main properties: \textit{BucketName} (Line 5-6) for name of the S3 bucket, \textit{cred\_file\_path} (Line 7-8) for path the file that contains login credentials, and \textit{Region} (Line 9-10) of the S3 bucket, as shown in Listing \ref{code:ConsS3Bucket_properties}. On the contrary to the remote data consumption, \textit{ConsumeLocal} TOSCA node type is developed that would allow the user to consume the data from the local file system. The local file system refers to the file system where the pipeline is deployed. 

\begin{lstlisting}[language=xml, label=code:ConsGCSBucket_properties caption= Properties of \textit{ConsGCSBucket} TOSCA node type.]
node_types:
  radon.nodes.datapipeline.source.ConsGCSBucket:
    derived_from: radon.nodes.datapipeline.source.ConsumeRemote
    properties:
      bucket:
        type: string
        description: Name of the bucket
        required: false
      project_ID:
        type: string
        description: ID of the project.
      credential_JSON_file:
        type: string
        description: Path of the credentials JSON file
\end{lstlisting}       

\begin{lstlisting}[language=xml, label=code:ConsS3Bucket_properties, caption= Properties of \textit{ConsS3Bucket} TOSCA node type.]
node_types:
  radon.nodes.datapipeline.source.ConsS3Bucket:
    derived_from: radon.nodes.datapipeline.source.ConsumeRemote
    properties:
      BucketName:
        type: string
      cred_file_path:
        type: string
      Region:
        type: string
\end{lstlisting}

It is to be observed that all the TOSCA node types derived from SourcePB are based on NiFi technology, and hence in the implementation of each such node type may contain one or more NiFi processors.

\subsection{MidwayPB}
This subsection discusses the category of TOSCA data pipeline node types specifically developed to process the data while migrating from one system to another. This set of node types allows the developers to transform the data either locally or by invoking remote serverless functions.

Processing the consumed data before publishing it is another basic building block to develop data flow-based cloud applications. To fulfill such requirements, \textit{MidwayPB}\footnote{\url{https://github.com/chinmaya-dehury/radon-particles/nodetypes/radon.nodes.datapipeline.process}} TOSCA node type is developed. \textit{MidwayPB} TOSCA node type is developed to provide a parent node type for all potential data processing related TOSCA node types. \textit{MidwayPB} TOSCA node type is derived from \textit{PipelineBlock}. 

{
    Similar to \textit{SourcePB} TOSCA node type, \textit{MidwayPB} (in Listing \ref{code:midwaypb_requirements}) also has three requirements: (a) \textit{connectToPipeline}: requirement to connect to a local pipeline, in Line 6-10, (b) \textit{connectToPipelineRemote}: requirement to connect to a remote pipeline, in Line 16-20, and (c) \textit{host}: requirement of an hosting environment, in Line 11-15. The requirements of \textit{MidwayPB} and \textit{SourcePB} TOSCA node type have same purpose. 
}
{
However, unlike \textit{SourcePB}, \textit{MidwayPB} has the capabilities to receive the data from other local pipelines (\textit{ConnectToPipeline}, in Line 26-29) and remote pipelines (\textit{ConnectToPipelineRemote}, in Line 22-25), as shown in Listing \ref{code:midwaypb_requirements}. Both the capabilities are of types \textit{radon.capabilities.datapipeline.ConnectToPipeline} and have the valid source type of \textit{SourcePB} and \textit{MidwayPB}. It is possible that the data can be consumed from multiple data sources using multiple \textit{SourcePB} type TOSCA nodes. Moreover, the output of one \textit{MidwayPB} type TOSCA node can be input to other node of type \textit{MidwayPB}. To facilitate such feature, the upper bound of receiving data from other local and remote pipelines is set to \textit{UNBOUNDED}, as shown in Listing \ref{code:midwaypb_requirements}. }
\begin{lstlisting}[language=xml, label=code:midwaypb_requirements, caption= List of requirements and capabilities for \textit{MidwayPB} TOSCA node type.]
tosca_definitions_version: tosca_simple_yaml_1_3
node_types:
  radon.nodes.datapipeline.MidwayPB:
    derived_from: radon.nodes.datapipeline.PipelineBlock
    requirements:
      - ConnectToPipeline:
          capability: radon.capabilities.datapipeline.ConnectToPipeline
          node: radon.nodes.datapipeline.PipelineBlock
          relationship: radon.relationships.datapipeline.ConnectNifiLocal
          occurrences: [ 1, UNBOUNDED ]
      - host:
          capability: tosca.capabilities.Container
          node: radon.nodes.nifi.Nifi
          relationship: tosca.relationships.HostedOn
          occurrences: [ 1, 1 ]
      - ConnectToPipelineRemote:
          capability: radon.capabilities.datapipeline.ConnectToPipeline
          node: radon.nodes.datapipeline.PipelineBlock
          relationship: radon.relationships.datapipeline.ConnectNifiRemote
          occurrences: [ 1, UNBOUNDED ]
    capabilities:
      ConnectToPipelineRemote:
        occurrences: [ 1, UNBOUNDED ]
        valid_source_types: [ radon.nodes.datapipeline.SourcePB, radon.nodes.datapipeline.MidwayPB ]
        type: radon.capabilities.datapipeline.ConnectToPipeline
      ConnectToPipeline:
        occurrences: [ 1, UNBOUNDED ]
        valid_source_types: [ radon.nodes.datapipeline.SourcePB, radon.nodes.datapipeline.MidwayPB ]
        type: radon.capabilities.datapipeline.ConnectToPipeline
\end{lstlisting}

{
Based on the processing location and the functionality, three different node types are further created: \textit{LocalAction} with the purpose to process the data on the local machine, \textit{RemoteAction} to invoke a remote serverless function to process the data, and \textit{RouteToRemote} to route the data to other local or remote pipelines based on the specific condition. Handling the data on the local machine is suitable in case of data fusion, data integration, data cleaning, etc. Further, this is suitable for applying analytic tasks on large scale data, where invoking the serverless function would not give the desired performance due to the size of the task or due to the nature of the data analysis job. To provide such functionalities, a number of TOSCA node types are developed that are derived from \textit{LocalAction} TOSCA node type. \textit{ExecuteCommand}, \textit{ExecutePython}, and \textit{ExecuteRuby} TOSCA nodes, as shown in Figure \ref{fig:ModelHrchy}, are developed allowing the user to invoke and execute the data analytics task in shell terminal, Python engine, and in Ruby engine, respectively. With this, users need to provide the path to the scripting code or provide the scripting code to process and analyze the input data. Moreover, user can also encrypt the raw data with \textit{Encrypt} TOSCA node and decrypt the encrypted data using \textit{Decrypt} TOSCA node on the local machine. 

The modern data-flow based cloud applications rely highly on a serverless platform as well, which not only allows the developer to detach themselves from infrastructure and platform management responsibilities, including resource provisioning, scaling, and maintenance, but also minimize the service cost and application development time. To fulfill this requirement of such modern serverless cloud applications, a number of TOSCA node types are developed that are derived from \textit{RemoteAction}. The developed \textit{InvokeLambda} and \textit{InvokeOpenFaaS} TOSCA node types can be used to invoke AWS Lambda serverless function and OpenFaaS serverless function, respectively. While invoking such remote serverless function, the received data are sent in JSON format as a part of a body of the HTTP request. 
In general, the route of the data is mostly decided by the developer in the design time of the cloud application. In some exceptional situations, the data can be routed based on conditions defined in dedicated data routing programs installed on local machines. However, to make the data routing job more dynamic, we have envisioned to develop a TOSCA node type, \textit{RouteToRemote}, that handles routing of data in a dynamic manner. Dynamic routing refers to deciding the next destination of data on its arrival. The next destination can be another \textit{MidwayPB} TOSCA node or a TOSCA node to publish the data. Apache NiFi is used as the underlined open-source data management tool to develop all the above TOSCA node types that are derived from \textit{MidwayPB}. 
}

\subsection{DestinationPB}
This subsection discusses the developed TOSCA node types that are used to publish the data to external storage systems such as FTP server, Google storage bucket, AWS S3 bucket, MinIO servers, MQTT broker, Azure storage system etc.

{ 
\textit{DestinationPB}\footnote{\url{https://github.com/radon-h2020/radon-particles/nodetypes/radon.nodes.datapipeline.destination}} TOSCA node type, as shown in Listing \ref{code:destpb_requirements}, is created as the counterpart of \textit{SourcePB} TOSCA node type. Unlike \textit{SourcePB}, \textit{DestinationPB} mainly focuses on publishing the data/result to local or remote storage server provided by private or public cloud provider. As shown in Figure \ref{fig:BasicBuildingBlocks}, multiple nodes in TOSCA service template that are of type \textit{SourcePB} and \textit{DestinationPB} may refer same storage location as both data source and data destination. The storage server can be located in the local machine or in the remote machine. For this purpose, \textit{PublishRemote} and \textit{PublishLocal} TOSCA node types are developed that are derived from \textit{DestinationPB}. For each TOSCA node type developed under \textit{SourcePB}, the corresponding counterpart TOSCA node type is developed under \textit{DestinationPB}, which can be seen in Figure \ref{fig:ModelHrchy}.}

All the TOSCA node types that are derived \textit{DestinationPB}, as in Figure \ref{fig:ModelHrchy}, have the capabilities to accept the connections from other pipelines of type either \textit{SourcePB} or \textit{MidwayPB} and not from the nodes of the same type. The requirement and the capability \textit{DestinationPB} node type is given in Listing \ref{code:destpb_requirements}.

\begin{lstlisting}[language=xml, label=code:destpb_requirements, caption= List of requirements and capabilities for \textit{DestinationPB} TOSCA node type.]

tosca_definitions_version: tosca_simple_yaml_1_3
node_types:
  radon.nodes.datapipeline.DestinationPB:
    derived_from: radon.nodes.datapipeline.PipelineBlock
    requirements:
      - host:
          capability: tosca.capabilities.Container
          node: radon.nodes.nifi.Nifi
          relationship: tosca.relationships.HostedOn
          occurrences: [ 1, 1 ]
    capabilities:
      ConnectToPipelineRemote:
        occurrences: [ 1, UNBOUNDED ]
        valid_source_types: [ radon.nodes.datapipeline.SourcePB, radon.nodes.datapipeline.MidwayPB ]
        type: radon.capabilities.datapipeline.ConnectToPipeline
      ConnectToPipeline:
        occurrences: [ 1, UNBOUNDED ]
        valid_source_types: [ radon.nodes.datapipeline.MidwayPB, radon.nodes.datapipeline.SourcePB ]
        type: radon.capabilities.datapipeline.ConnectToPipeline
\end{lstlisting}


\subsection{Standalone}
This section discusses another category of the developed TOSCA node types that act as standalone data pipelines and do not need any input/output from/to another pipeline block.

Along with the basic building blocks discussed above, another building block, \textit{Standalone}\footnote{\url{https://github.com/chinmaya-dehury/radon-particles/tree/dp_tmplt/nodetypes/radon.nodes.datapipeline.Standalone}}, is created focusing on performing a very specific task, such as synchronising two AWS S3 buckets, taking data backup from AWS DynamoDB to AWS S3 bucket, etc., as shown in Figure \ref{fig:ModelHrchy}. Each TOSCA node type under the Standalone pipeline block can be very vendor-specific. In other words, these TOSCA node types focus on the flow and processing of data using the storage and computational service of a single cloud provider. The \textit{Standalone} TOSCA node type does not have any specific requirement or capabilities. Requirement and capabilities vary with the derived node types. As the name suggests, these node types don't send any data to other pipelines, nor do they have any capabilities to receive data from other pipelines. Such node types are developed by combining all the three basic building blocks. For example, the TOSCA node type, \textit{AWSCopyS3ToS3}, as in Figure \ref{fig:ModelHrchy}, copies the data from one S3 bucket to another S3 bucket by combining the functionalities of \textit{SourcePB} and \textit{DestinationPB} node types.

We have developed a number of standalone TOSCA data pipeline node types using the services provided by Amazon cloud. \textit{AWSCopyS3ToS3} node type is developed to synchronise two S3 buckets. This node type requires \textit{AWSPlatform} as a hosting platform. Some of the basic properties that the users need to provide are the name of the source and destination S3 bucket, the credential information, the S3 bucket name to store the logs, etc. It is also possible that the source and destination S3 bucket names are the same. In such a scenario, the source and destination directories need to be different. Similarly, \textit{AWSCopyDynamodbToS3} and \textit{AWSCopyS3ToDynamodb} TOSCA node types are developed to copy the data between Amazon cloud provided DynamoDB and S3 buckets. \textit{AWSShellCommand} TOSCA node type is developed to execute a command or a script. Similarly, to execute an SQL command on a database, \textit{AWSSqlActivity} TOSCA node type is developed. All the above-mentioned AWS-related TOSCA node types required \textit{AWSPlatform} as a hosting platform. These nodes can not be hosted on a private cloud or any other public cloud instances.

The above TOSCA node types provide the basic functionalities to handle smooth flow and transformation of the data by incorporating multiple cloud provider and their serverless platform. To realize the applicability and the advantage of TOSCAdata, the developed TOSCA node types are used to develop a cloud application, as discussed in a further section. 

\section{Application of TOSCAdata}\label{sec:usecase}
In previous sections, the proposed TOSCAdata is explained, including the detailed description of the TOSCA node types, requirements, capabilities, relationships, etc. TOSCAdata provides a number of necessary TOSCA node type definitions for building data pipeline-based cloud applications. It is to be noted that no additional orchestration or modelling tool is required other than the tools that provide the support for TOSCA to model a data pipeline-based cloud service. The steps required to develop any TOSCA-based cloud application can be followed to develop a data pipeline application for all the stakeholders. To demonstrate the functionalities, the proposed TOSCAdata is implemented in Viarota, a real-world mobile and web-based tourism promotion cloud application currently under development at Athens Technology Center (ATC), Greece \cite{viarota}. In addition to this use case, we also have demonstrated the movement of image data across four different cloud providers through \textit{image data migration and processing} example.

\subsection{Use case: Viarota application} 
In this section, a web-based tourism promotion cloud application is considered as a use case scenario and discussed how the proposed data pipeline-based TOSCA node types can be used to efficiently handle the data flow and transformation without much worrying about the vendor lock-in issue while migrating data from one cloud to other. 

\begin{figure*}[h]
    \centering
    \begin{subfigure}{0.9\textwidth}
        \includegraphics[width=1\textwidth]{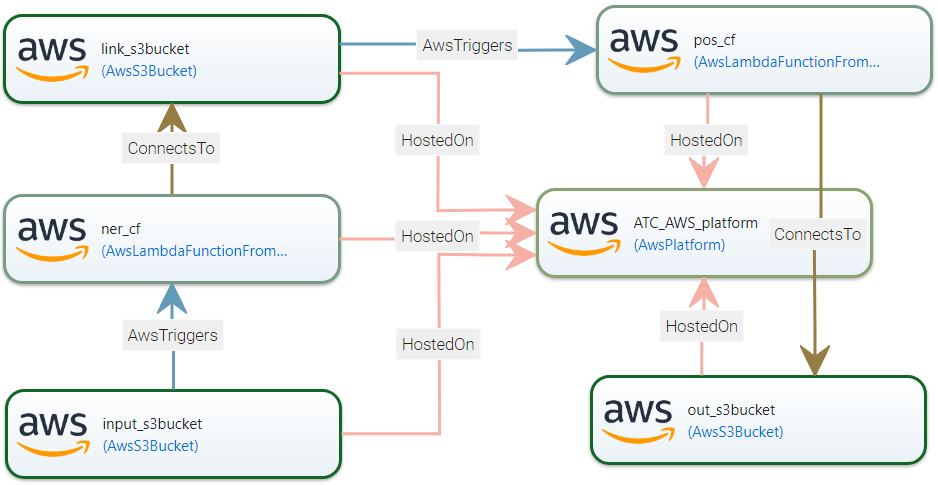}
        \caption{First part of the whole Viarota NLP pipeline.}
        \label{fig:viarotaNLPArchitecture:a}
    \end{subfigure} 
    \hfil
    \begin{subfigure}{0.9\textwidth}
        \includegraphics[width=1\textwidth]{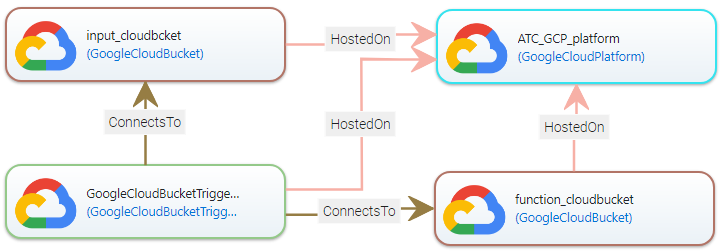}
        \caption{Second part of the whole Viarota NLP pipeline.}
        \label{fig:viarotaNLPArchitecture:b}
    \end{subfigure} 
    \hfil
    \begin{subfigure}{0.9\textwidth}
        \includegraphics[width=1\textwidth]{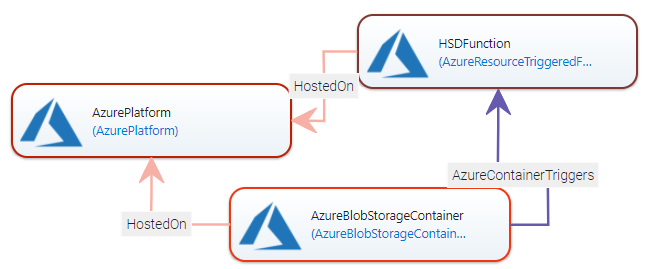}
        \caption{Third part of the whole Viarota NLP pipeline.}
        \label{fig:viarotaNLPArchitecture:c}
    \end{subfigure}
    \caption{Viarota NLP pipeline topology.}
    \label{fig:viarotaNLPArchitecture}
\end{figure*}

Viarota \cite{viarota} is a mobile as well as a web-based cloud application in the context of tourism promotion. Viarota enhances a visitor's travel experience by providing optimal loyalty-based personalized city tour planning, developed within the RADON project \cite{casale_radon_2020}. A tour plan is composed of Point of Interests (POIs) that follow the taxonomy concepts for the places of a visit being included in the tours proposed for a touristic destination. Viarota crawls related content from different social media sources like Twitter and YouTube as well as from tour related website RSS feeds. The crawled data are then processed and stored in a database for providing aggregated reviews on the visits placed in the proposed tours. The raw text, extracted from the crawled social media items, processing consists of a set of Natural Language Processing (NLP) functions, a sentiment analysis function and a hate speech detection function. Each function is modeled and implemented as a FaaS in the cloud platforms used, namely the Amazon Web Service (AWS), the Google Cloud Platform (GCP) and the Azure platform. To synchronise the data between these two different cloud storages, the data pipeline approach can be applied by providing a data link between the NLP/ML functions running on different cloud platforms as a single processing layer. 

\begin{figure*}[h]
 \centering
 \includegraphics[width=\textwidth]{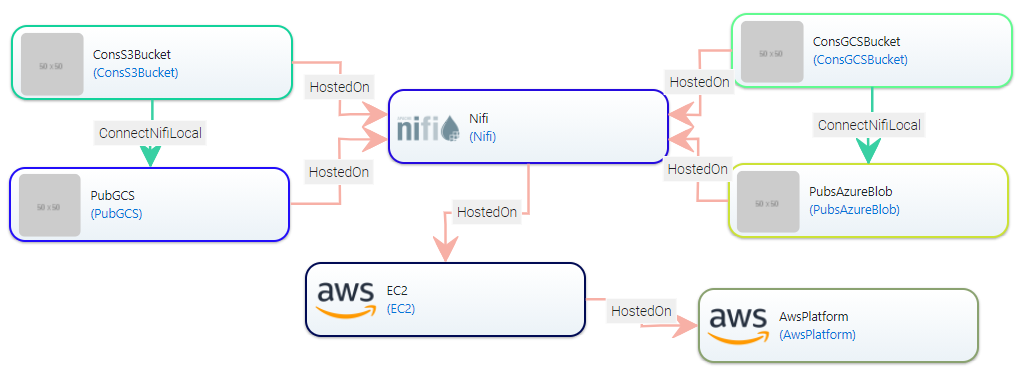}
 \caption{Viarota NLP data pipeline.}
 \label{fig:viarotaDataPipline}
\end{figure*}
 
The topology of the Viarota NLP/ML pipeline is modeled using Eclipse Winery, a graphical modelling tool for TOSCA-based cloud application \cite{kopp_winerymodeling_2013, eclipse_winery}, following the serverless execution model where the Named-entity recognition and the Part-of-speech tagging functions are hosted on AWS Lambda, Amazon’s FaaS platform, as shown in Figure \ref{fig:viarotaNLPArchitecture:a}, the sentiment analysis function is hosted on GCP Cloud Function, Google's FaaS platform, as shown in Figure \ref{fig:viarotaNLPArchitecture:b} and the hate speech detection function is hosted on Azure Functions, Azure's FaaS platform, as shown in Figure \ref{fig:viarotaNLPArchitecture:c}. The function deployment is modeled using the corresponding TOSCA node type definitions developed in the context of the RADON project. Following an event-driven programming model, cloud storage services are used to trigger the Viarota NLP/ML functions, acting as event sources. The Viarota social media crawlers continuously feed the NLP/ML functions with events in the form of new social media item insertions in the configured cloud storage input buckets. By configuring the output bucket of one NLP/ML function as the input bucket of another NLP/ML function, the social media items can traverse the whole processing pipeline without further intervention. However, the TOSCAdata approach is needed to be introduced in order to allow the stream of social media items and the corresponding analysis results to be transmitted between the different cloud platforms. 

The TOSCA data pipeline nodes allow establishing a multi-cloud integration layer that links the inputs and outputs of different cloud platforms. Specifically, the Viarota NLP functions (hosted on AWS) need to communicate the analysis results to the ML functions, the sentiment analysis (hosted on GCP) and the hate speech detection (hosted on Azure), in order to be taken into account in the processing tasks (i.e. words tagged as adjectives seems to capture strong sentiment polarity). There is the need to configure two cross-cloud data flows that would link the respective cloud storage of each platform, allowing the synchronization of data without further interventions. The first data flow (AWS - GCP) is modelled using the TOSCAdata nodes \emph{ConsS3Bucket} and \emph{PubsGCS} while the second data flow configuration (GCP - Azure) consists of the TOSCAdata nodes \emph{ConsGCSBucket} and \emph{PubsAzureBlob}, as shown in figure \ref{fig:viarotaDataPipline}. The TOSCAdata nodes modelled refer to the storage buckets configured in the Viarota pipeline topology (Figure \ref{fig:viarotaNLPArchitecture}). The node \emph{ConsS3Bucket} consumes events from an AWS S3 bucket and in conjunction with the \emph{PubsGCS} node, it replicates the storage events in a GCP cloud storage bucket while the \emph{ConsGCSBucket} node consumes from a GCP cloud bucket and it replicates the events to Azure blob storage (through node \emph{PubsAzureBlob}). 

This way, it is possible to synchronize data flows across different vendor storage buckets that would otherwise require to invoke vendor-specific technologies. Thus, the data pipeline layer that spans all cloud platforms (AWS, GCP and Azure) promotes data lock-in avoidance in the architecture of the Viarota application. Although, existing solutions such as Apache Airflow\footnote{https://airflow.apache.org/} and Node-RED\footnote{https://nodered.org/}, can address the data lock-in issue, the fact that the TOSCAdata tool is considered advantageous since it provides native integration with the Viarota models that are also developed based on TOSCA language. The TOSCA specification addresses the lack of standardisation and contributes to the portability of the implemented Viarota functions. Moreover, the Apache Airflow tool requires writing code for implementing tasks like storage replication, while the TOSCAdata tool allows to drag and drop the YAML based data pipeline node definitions in a more declarative and user-friendly way while designing the topology.

\subsection{Image data migration and processing application} \label{sec:usecase:myDemo}

\begin{figure*}[h]
 \centering
 \includegraphics[width=0.95\linewidth]{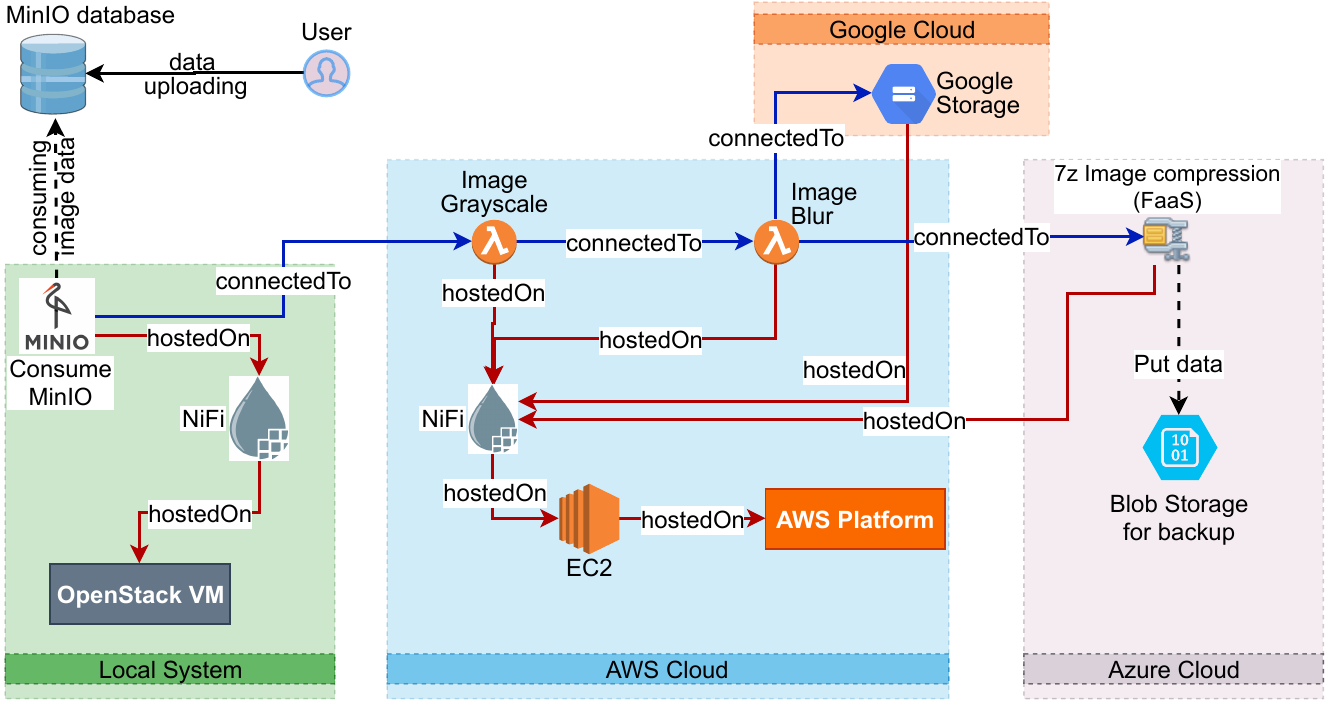}
 \caption{Image data migration and processing using TOSCAdata.}
 \label{fig:dp_demo}
\end{figure*}

\begin{figure*}[h]
 \centering
 \includegraphics[width=0.95\linewidth]{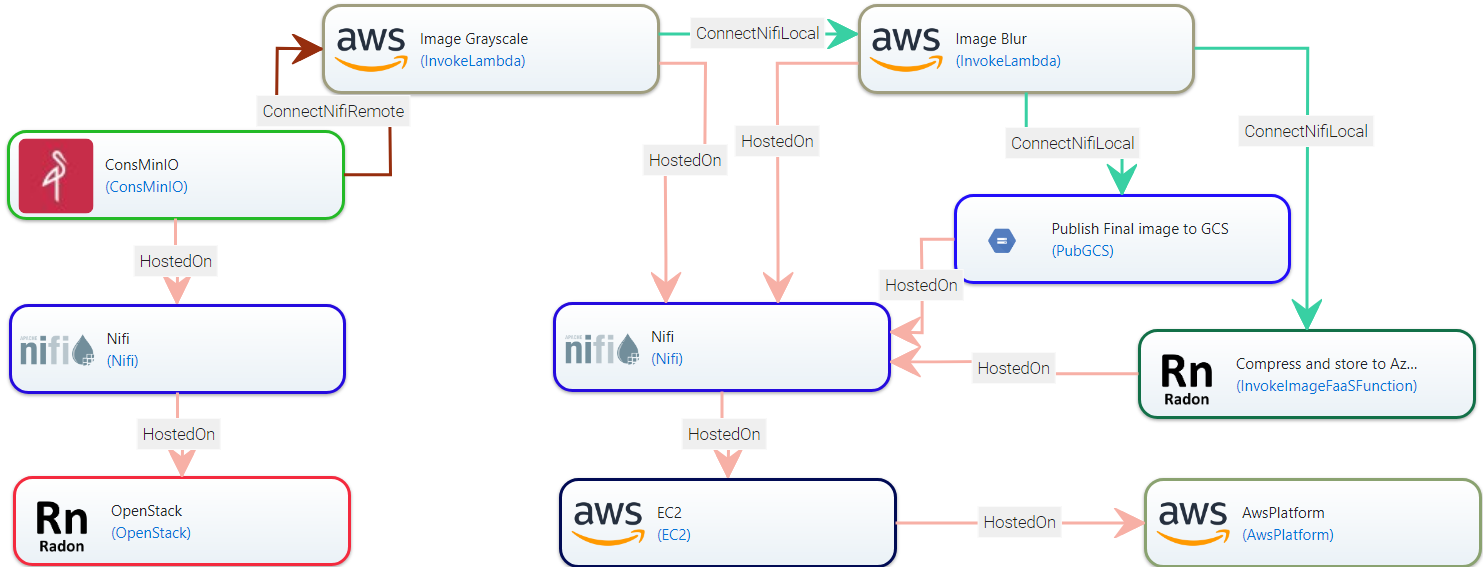}
 \caption{Modelling of data pipeline example (given in Figure 8) using RADON graphical modelling tool.}
 \label{fig:dp_demo_gmt}
\end{figure*}

In addition to the above implementation of a real-world cloud application, an \textit{image data migration and processing} application is implemented using TOSCAdata. Figure \ref{fig:dp_demo} depicts the flow of data from an on-premise MinIO database server to OpenStack private cloud, Google cloud storage, and Azure storage system. As shown in Figure \ref{fig:dp_demo}, four cloud environments are used: OpenStack private cloud, AWS cloud, Google cloud, and Azure cloud. The automatic data migration takes place upon manual uploading of the image data by the user to the local MinIO server. The uploaded data are automatically passed through OpenStack cloud, processed by AWS serverless platform, and then stored in Google cloud storage and compressed by Azure function, which eventually are stored in Azure blob storage.

Atop OpenStack VM, NiFi software component is installed and configured using \textit{NiFi} TOSCA node. Atop \textit{NiFi} node \textit{ConsMinIO}\footnote{\url{https://github.com/chinmaya-dehury/radon-particles/tree/dp_tmplt_part3/nodetypes/radon.nodes.datapipeline.source/ConsMinIO}} data pipeline node is deployed with the properties as mentioned in Listing \ref{code:demo:consminio}. This data pipeline node is used to consume the data from the local MinIO database server. 
\begin{lstlisting}[language=xml, label=code:demo:consminio, caption= \textit{ConsMinIO} data pipeline node.]
ConsMinIO_0:
  type: radon.nodes.datapipeline.source.ConsMinIO
  properties:
    BucketName: "firstbucket"
    cred_file_path: "{ get_artifact: [SELF, credentials]}"
    MinIO_Endpoint: "http://172.17.25.36:8089"
    schedulingStrategy: "EVENT_DRIVEN"
    schedulingPeriodCRON: "* * * * * ?"
\end{lstlisting}

The \textit{ConsMinIO} node forwards the image data to the \textit{InvokeLambda}\footnote{\url{https://github.com/chinmaya-dehury/radon-particles/tree/dp_tmplt_part3/nodetypes/radon.nodes.datapipeline.process/InvokeLambda}} data pipeline node, which invokes an image processing function deployed on AWS lambda serverless platform to grayscale the original image. The properties of this node are provided in Listing \ref{code:demo:InvokeLambda1}. The grayscaled image is further forwarded to another \textit{InvokeLambda} data pipeline node that invokes an image processing Lambda function to blur the grayscaled image. The properties of this node are provided in Listing \ref{code:demo:InvokeLambda2}. Both the \textit{InvokeLambda} data pipeline nodes are deployed on a NiFi software component hosted on AWS EC2 instance, as shown in \ref{fig:dp_demo} and \ref{fig:dp_demo_gmt}. 

\begin{lstlisting}[language=xml, label=code:demo:InvokeLambda1, caption= Properties of \textit{InvokeLambda} data pipeline node for image grayscale. ]
InvokeLambda_0:
  type: radon.nodes.datapipeline.process.InvokeLambda
  properties:
    cred_file_path: "{ get_artifact: [SELF, credFile]}"
    schedulingStrategy: "EVENT_DRIVEN"
    function_name: "img-grayscale-nifi"
    schedulingPeriodCRON: "* * * * * ?"
    region: "eu-west-1"
\end{lstlisting}

\begin{lstlisting}[language=xml, label=code:demo:InvokeLambda2, caption= Properties of \textit{InvokeLambda} data pipeline node for image blurring.]
InvokeLambda_1:
  type: radon.nodes.datapipeline.process.InvokeLambda
  properties:
    cred_file_path: "{ get_artifact: [SELF, credFile]}"
    schedulingStrategy: "EVENT_DRIVEN"
    function_name: "img-blur-nifi"
    schedulingPeriodCRON: "* * * * * ?"
    region: "eu-west-1"
\end{lstlisting}

The second \textit{InvokeLambda} data pipeline node forwards the blurred image to two data pipeline nodes: \textit{PubGCS}\footnote{\url{https://github.com/chinmaya-dehury/radon-particles/tree/dp_tmplt_part3/nodetypes/radon.nodes.datapipeline.destination/PubGCS}} and \textit{InvokeImageFaaSFunction}\footnote{\url{https://github.com/chinmaya-dehury/radon-particles/tree/dp_tmplt_part3/nodetypes/radon.nodes.datapipeline.process/InvokeImageFaaSFunction}}. The \textit{PubGCS} data pipeline node (configured with the properties shown in Listing \ref{code:demo:PubGCS}) store the blurred image to Google Cloud Storage. On the other hand, the \textit{InvokeImageFaaSFunction} data pipeline node is used to invoke a function with image data as arguments deployed on the Azure serverless platform. The \textit{InvokeImageFaaSFunction} data pipeline node is configured with the properties given in Listing \ref{code:demo:InvokeImageFaaSFunction}. The Azure function is developed to compress the received blurred image and store it in an Azure storage container, as shown in Figure \ref{fig:dp_demo}. The detailed source code of this implementation can be found in GitHub repository \cite{dp-webinar}.  
\begin{lstlisting}[language=xml,label=code:demo:PubGCS, caption= Properties of \textit{PubGCS} data pipeline node for storing the blur image to Google storage.]
PubGCS_0:
  type: radon.nodes.datapipeline.destination.PubGCS
  properties:
    BucketName: "radongcs"
    cred_file_path: "{ get_artifact: [SELF, credFile ] }"
    schedulingStrategy: "EVENT_DRIVEN"
    ProjectID: "radon-825040-utr"
    schedulingPeriodCRON: "* * * * * ?"
\end{lstlisting}

\begin{lstlisting}[language=xml, label=code:demo:InvokeImageFaaSFunction, caption= Properties of \textit{InvokeImageFaaSFunction} data pipeline node.]
InvokeImageFaaSFunction_0:
  type: radon.nodes.datapipeline.process.InvokeImageFaaSFunction
  properties:
    function_URL: "AWS_function_endpoint_here"
    schedulingStrategy: "EVENT_DRIVEN"
    schedulingPeriodCRON: "* * * * * ?"
    HTTP_method: "POST"
\end{lstlisting}


\subsection{Discussion}
In the above sections, we have demonstrated how to apply the TOSCAdata on real-world use cases to implement real-time data migration services across multi-cloud and on-premise environments. In the case of the Viarota application, it enabled real time data synchronization between AWS, Azure and Google Cloud, supporting a serverless application where FaaS functions are deployed strategically in different clouds. In the case of image processing application, we also demonstrated how to integrate on-premise systems deployed in OpenStack, how to chain Serverless functions and how to utilize data pipelines deployed across multiple NiFi deployments.

As a result of providing TOSCA data pipeline blocks that users can simply drag and drop into a TOSCA graphical modelling tool/editor, it reduces the development effort of designing data integration and processing services. As a result of each data pipeline block being designed as an individual software service with its own life cycle control scripts (implemented in Ansible), TOSCAdata enables the data-pipelines-as-code pattern, allowing the automated deployment of data pipelines. For example, as part of CI/CD pipelines. TOSCAdata also reduces the effect of data lock-in when dealing with data migration and processing service, as the data pipeline blocks can be deployed on any of the supported public cloud providers (e.g AWS, Azure, Google Cloud) and other platforms like OpenStack. This approach also hides any low level, platform specific complexities of designing multi cloud data synchronisation services.

\section{Conclusions and future works}\label{sec:concl}
{
TOSCA standard focuses on portability and interoperability of the cloud application through a service blueprint that organizes the service components and the relationship among themselves in a graph structure. In this work, we have extended the capability of TOSCA and introduces TOSCAdata, focusing on the modelling of data pipeline-based cloud application. TOSCAdata not only provides a set of TOSCA models for designing a provider and technology-specific cloud application but also empowers the developers to integrate the multiple services provided by multiple cloud providers, different serverless platform, and microservices without compromising the smooth flow and transformation of data. With the developed TOSCA models, it becomes easy to migrate the data across multiple private and public cloud environments through rapid modelling, development, and deployment, leveraging the capabilities of modern data-intensive cloud applications, as discussed in Section \ref{sec:cap_features}. TOSCAdata is developed to provide a unified solution to the software development community that works atop one open-source (Apache NiFi) and one commercial data management solution (AWS Datapipeline), which further can be extended by providing the support for a number of other commercial data management platforms. }

It should be noted that the current development of TOSCAdata does not provide all possible TOSCA models that could be essential to build a data pipeline-based cloud applications, and hence there are rooms for its further development. Currently, TOSCAdata supports one open-source and one commercial data management solutions. The further development to support more number of commercial data management platform from different cloud providers is a part of our future work. Scalability and a higher level of security of data is another development direction that we would like to improve TOSCAdata. Further, our future works will also include the realization of the performance of TOSCA data pipeline-based cloud applications by experimenting with a large scale multi-cloud environment.

\section*{Acknowledgment}
This work is partially funded by the European Union's Horizon 2020 research and innovation project RADON (825040). We also thank financial support to UoH-IoE by MHRD (F11/9/2019-U3(A)).

\bibliographystyle{elsarticle-num}
\bibliography{referencefile} 

\par \noindent
\begin{wrapfigure}{i}{1in}
\includegraphics[width=1in,keepaspectratio]{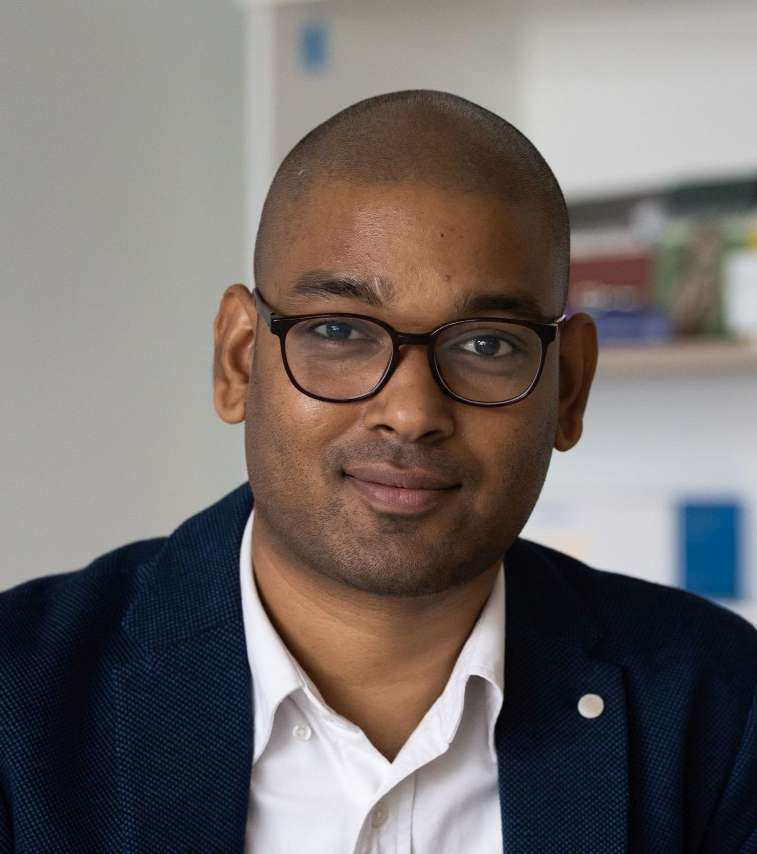}
\end{wrapfigure}
\textbf{Chinmaya Kumar Dehury} received bachelor degree from Sambalpur University, India, in June 2009 and MCA degree from Biju Pattnaik University of Technology, India, in June 2013. He received the PhD Degree in the department of Computer Science and Information Engineering, Chang Gung University, Taiwan. Currently, he is a postdoctoral research fellow in the Mobile \& Cloud Lab, Institute of Computer Science, University of Tartu, Estonia. His research interests include scheduling, resource management and fault tolerance problems of Cloud and fog Computing, and the application of artificial intelligence in cloud management. He is an reviewer to several journals and conferences, such as IEEE TPDS, IEEE JSAC, Wiley Software: Practice and Experience, etc.

\par \noindent
\begin{wrapfigure}{i}{1in}
\includegraphics[width=1in,keepaspectratio]{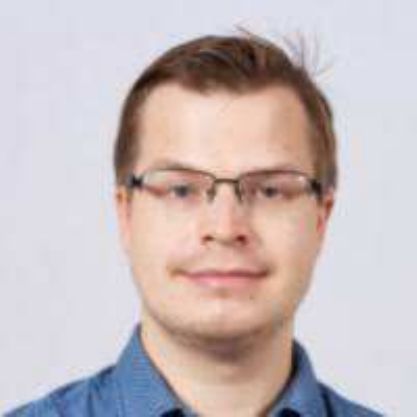}
\end{wrapfigure}
\textbf{Pelle Jakovits} received his Ph.D. in computer science from University of Tartu in March 2017 on topic "Adapting Scientific Computing Algorithms to Distributed Computing Frameworks". His main research interests are algorithm parallelization, high-performance computing in the cloud and efficiency of real-time stream data analytics. He has contributed to several EU funded projects, such as H2020 RADON (http://radon-h2020.eu/), FP7 REMICS (http://remics.eu/) and I4W Robo M.D. He has significant experience in setting up various scale distributed computing environments in the cloud from Hadoop (Cloudera CDH or custom) ecosystems to IoT platforms like Cumulicity using OpenStack, Docker, Ansible, Chef and shell scripts. He is teaching courses related to cloud computing and large scale data processing.

\par \noindent
\begin{wrapfigure}{i}{1in}\includegraphics[width=1in,keepaspectratio]{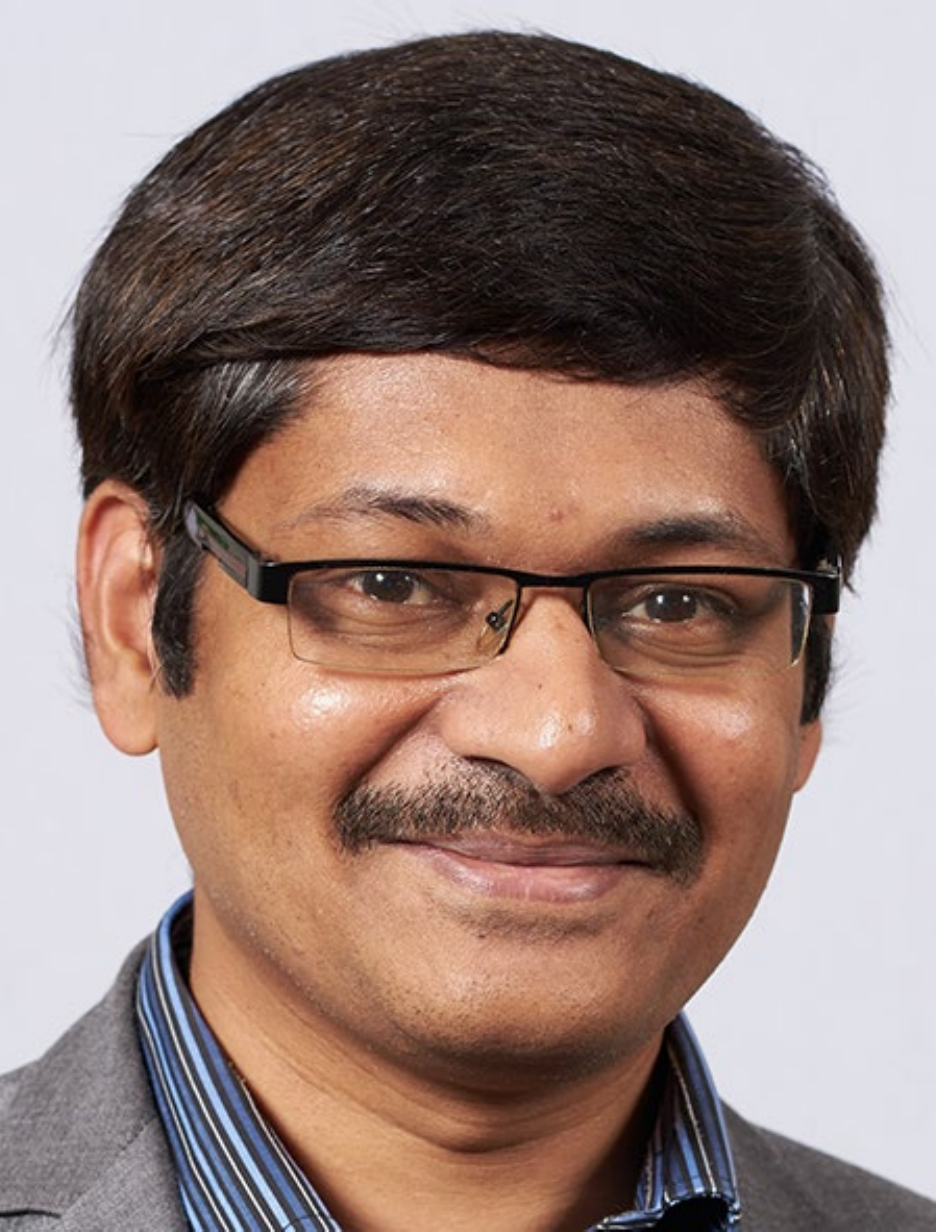}\end{wrapfigure}\textbf{Satish Narayana Srirama} is an Associate Professor at the School of Computer and Information Sciences, University of Hyderabad, India. He is also a Visiting Professor and the honorary head of the Mobile \& Cloud Lab at the Institute of Computer Science, University of Tartu, Estonia, which he led as a Research Professor until June 2020. He received his PhD in computer science from RWTH Aachen University, Germany in 2008. His current research focuses on cloud computing, mobile web services, mobile cloud, Internet of Things, fog computing, migrating scientific computing and enterprise applications to the cloud and large-scale data analytics on the cloud. He is IEEE Senior Member, an Editor of Wiley Software: Practice and Experience, a 50 year old Journal, was an Associate Editor of IEEE Transactions in Cloud Computing and a program committee member of several international conferences and workshops. Dr. Srirama has co-authored over 150 refereed scientific publications in international conferences and journals. For further information of Prof. Srirama, please visit: http://kodu.ut.ee/$\backsim$srirama/.

\par \noindent
\begin{wrapfigure}{i}{1in}
\includegraphics[width=1in,keepaspectratio]{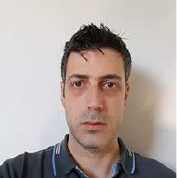}
\end{wrapfigure}
\textbf{Giorgos Giotis} Giotis holds a BSc in Informatics and Telecommunications from the National and Kapodistrian University of Athens and a MSc in Information Systems from Athens University of Economics and Business. He has been working as a software engineer in the R\&D department of ATC since 2011. His research interests include data mining, big data analytics and distributed applications.

\par \noindent
\begin{wrapfigure}{i}{1in}
\includegraphics[width=1in,keepaspectratio]{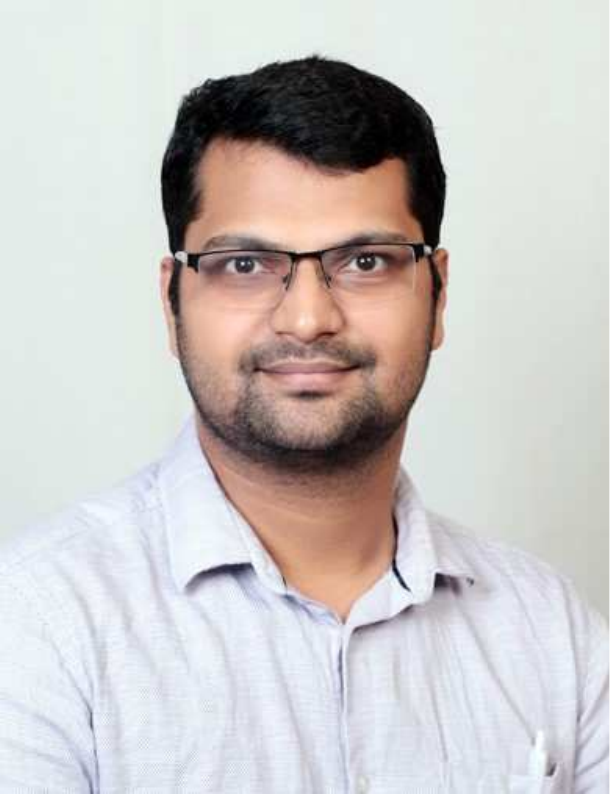}
\end{wrapfigure}
\textbf{Gaurav Garg} received a bachelor's degree from Punjab Technical University, India, in June 2011. He has worked as an application engineer with FANUC for more than 4 years, where he worked on IoT and control system projects. Currently, he is pursuing a Master's degree in Robotics and Computer Engineering and working as a research assistant in the Mobile \& Cloud Lab, Institute of Computer Science, University of Tartu, Estonia.
\end{document}